\documentclass[notitlepage]{article}

\def\BibTeX{{\rm B\kern-.05em{\sc i\kern-.025em b}\kern-.08emT\kern-.1667em\lower.7ex\hbox{E}\kern-.125emX}}

\PassOptionsToPackage{hyphens}{url}\usepackage{hyperref}

\usepackage{subfigure}
\usepackage{graphicx}
\usepackage{booktabs, footnote}
\usepackage{enumitem}
\usepackage{longtable}
\usepackage{amsfonts}
\usepackage[numbers]{natbib}
\usepackage{amsmath}
\usepackage{graphics}
\usepackage[showframe=false]{geometry}
\usepackage{changepage}
\usepackage{parskip}
\usepackage{authblk}
\usepackage[utf8]{inputenc} 

\providecommand{\keywords}[1]
{
  \small	
  \textbf{\textit{Keywords---}} #1
}

\title{Comparison and Benchmark of Graph Clustering Algorithms}

\author{Lizhen Shi$^{1}$, Bo Chen$^{2}$  \\
        \small $^{1}$Department of Computer Science, Florida State University, Tallahassee, FL, USA\\
        \small $^{2}$bochen0909@gmail.com\\
}

\date{} 

\begin{document}

\maketitle

\begin{abstract}
Graph clustering is widely used in analysis of biological networks, social networks and etc. 
For over a decade many graph clustering algorithms have been published, however a comprehensive and consistent performance comparison is not available. 
In this paper we benchmarked more than 70 graph clustering programs to evaluate their runtime and quality performance for both weighted and unweighted graphs. We also analyzed the characteristics of ground truth that affects the performance.
Our work is capable to not only supply a start point for engineers to select clustering algorithms but also could provide a viewpoint for researchers to design new  algorithms. 
\end{abstract}


\section{Introduction}

Comparison of graph clustering algorithms is not new and has been discussed in academic literature, but not thoroughly. Most of comparisons are only for a few of them and seldom report runtime performance.
 \citep{Lancichinetti2009} leveraged LFR benchamrk to test clustering perfomrance for about 11 algorithms up to year 2010. 
\citep{Lancichinetti} tested community detection algorithms for overlapped graphs with CPM algorithm. \citep{yang2016comparative} benchmarked 8 algorithms  available in iGraph~\citep{igraph}.

When we were working on our previous work of Sparc~\citep{shi2019sparc,li2020deconvolute}, it was demanding to know a better algorithm than LPA that works on large graphs. However we can not find a comprehensive study which resulting initiating this work. In this paper we tested on more than 70 clustering programs published in last decade. We both analyzed their quality performance and runtime performance.

\subsection*{Definition}

A \textit{graph} $G$ is defined as a set of vertices $V$ and a set of edges $E$ where $E=\{(u,v)| u \in V, v \in V \}$. We use $m=|E|$ and $n=|V|$ to denote the number of edges and number of vertices.

An \textit{undirected graph} does not distinguish the order of the vertices of an edge  $(u,v)$, while a \textit{directed graph} takes $(u,v)$ and $(v,u)$ as two different edges. To be uniform we always put both $(u,v)$ and $(v,u)$ in $E$ for an undirected graph. If it is not specified in this text, it always assumes undirected graphs. 

A weighted function may defined on edges as $w: E \rightarrow \mathbb{R} $ to consider the importance of edges. Most commonly weights are always positive. If it is not specified in this text, it always assumes unweighted graphs. 

A \textit{clustering} $\mathcal{C}$ of a graph is a collection of subset of $V$, that is $\mathcal{C} = \{ c| c \subset V \}$. Each element in $\mathcal{C}$ is called a \textit{cluster} or a \textit{community}.  It usually assumes that the union of clusters of a clustering covers $V$ (otherwise we can always create a new cluster for the remaining vertices). If  $ \forall c_1 \in \mathcal{C},  \forall c_2 \in \mathcal{C}$, $c_1 \cap c_2 = \emptyset $, $\mathcal{C}$ is not overlapped, others it is \textit{overlapped}. 
If it is not specified in this text, it always assumes it is not overlapped. 

\subsection{Graph Properties}

Here is some commonly used properties of a graph.

\subsection*{Density}
The density of a graph is defined as 
 $ \frac{m}{n(n-1)} $

\subsection*{Diameter}
The \textit{diameter} of a graph is the length of its longest shortest path among all pairs of nodes.  
\textit{effect diameter} is more stable and defined as 90th of the distribution of shortest path lengths.

\subsection*{Centrality}
\textit{Degree centrality} of a vertex $v$ is defined as $\frac{\deg{(v)}}{(n-1)}$. 
\textit{Farness centrality} of a vertex $v$ is  the average shortest path length to all other nodes that reside in the same connected component as $v$. 
\textit{Farness centrality} is the  reciprocal of farness centrality.

\subsection*{Eccentricity}
\textit{Eccentricity} of a vertex $v$ is the largest shortest-path distance from the node to any other node in the Graph.

\subsection*{Clustering coefficient}

\textit{Global clustering coefficient} is defined as the ratio between the number of closed triplets and the number of all triplets. \textit{Local clustering coefficient} is defined in \citep{watts1998collective} to determine whether a graph is a small-world network.

\section{Clustering Measures}

To evaluate the quality of a clustering, both \textit{fitness measures}  and \textit{score measures} can be used. 

\subsection{Fitness Measures}
Fitness measures evaluate the performance of a clustering by inspecting the properties of intraconnection and interconnection of clusters. 
Fitness measures are defined on a graph or a subgraph. Many algorithms depend on certain measures to make a decision. Although they are vital in designing algorithms,  we found that they are not very useful when evaluating clustering result. However we put them here briefly for completeness. Please refer to \citep{Yang2012} which provides a more detailed description for some of them.

\subsection*{sum of intra weights}
The summation of intra weights for a cluster $c$ is defined as 
$$\text{sum\_intra\_weight}(c) = \sum_{v \in c, u \in c} w(v,u) $$ where  $w$ is weight function.

\subsection*{sum of out weights}
The summation of out weights for a cluster $c$ is defined as 
$$\text{sum\_out\_weight}(c) = \sum_{v \in c, u \notin c} w(v,u) $$ where  $w$ is weight function.

\subsection*{expansion}
\textit{Expansion} of a cluster $c$ is defined as 
$$ \textit{expansion}(c) = \frac{\text{sum\_out\_weight}(c)}{|c|}$$.

\subsection*{cut ratio}
\textit{Cut ratio} of a cluster $c$ is defined as 
$$\text{cut\_ratio}(c) = \frac{\text{sum\_out\_weight}(c)}{|c|(n-|c|)}$$.

\subsection*{intra cluster density}
\textit{Intra cluster density} of a cluster $c$ for a unweighted graph is defined as 
$$\text{intra\_cluster\_density}(c) = \frac{|E_c|}{|c|(|c|+1)}$$ where 
$E_c$ is the edge set of subgraph induced by $c$. 

\subsection*{inter cluster density}
\textit{Inter cluster density} of a cluster $c$ for a unweighted graph is defined as 
$$\text{inter\_cluster\_density}(c) = \frac{|\{ (u,v) | u \in c_i, v \in c_j, i \neq j, c_i \in \mathcal{C},c_j \in \mathcal{C}   \}|}{n(n-1) + \sum_{c \in \mathcal{C}}  |V_c|(n-|V_c|)} $$ where 
$V_c$ is the vertices set of subgraph induced by $c$. 

\subsection*{relative cluster density}
\textit{relative cluster density} of a cluster $c$  is defined as 
$$\text{relative\_cluster\_density}(c) = \frac{1}{1+ \frac{\text{sum\_out\_weight}(c) }{\text{sum\_intra\_weight}(c)}} $$.

\subsection*{modularity}
The modularity of a unweighted graph is defined as 
$$ \frac{\sum_{c \in \mathcal{C}} \text{sum\_intra\_weight}(c)}{A} -\frac{\sum_{v \in V} \text{sum\_degree}(v)^2}{A^2} $$ 
where $A=\sum_{v \in V} \deg(v)$ is the sum of weights.
For unweighted graph it is indentical to 
$$\frac{1}{m}  \sum_{i,j} (A_{ij}- \frac{k_i k_j}{m}) 1_{(i=j)}$$
where $A$ is the adjacency matrix ,  $1$ is indicator function, $k_i k_j$ is the expected number of random edges between the two nodes. 

\subsection*{conductance}
\textit{Conductance} of a cluster $c$  is defined as 
$$\text{conductance}(c) = \frac{\text{sum\_out\_weight}(c)}{\text{sum\_out\_weight}(c)+ \text{sum\_intra\_weight}(c)}  $$

\subsection*{normalized cut}
\textit{Normalized cut} of a cluster $c$  is defined as 
\begin{equation*}
\begin{split}
\text{normalized\_cut}(c) = \frac{\text{sum\_out\_weight}(c)}{\text{sum\_out\_weight}(c)+ \text{sum\_intra\_weight}(c)} \\
+\frac{\text{sum\_out\_weight}(c)}{\text{sum\_out\_weight}(c)+ +m-\text{sum\_intra\_weight}(c)} 
\end{split}
\end{equation*}

\subsection*{out degree fraction}

\textit{Maximum out degree fraction} of a cluster $c$  is defined as 
$$
\text{max\_ODF}(c) = \max_{u \in c} \frac{|\{ (u,v) | (u,v) \in E, v \notin c\}|}{\deg(u)}
$$

\textit{Average out degree fraction} of a cluster $c$  is defined as 
$$
\text{avg\_ODF}(c) = \frac{1}{\mathcal(C)} \sum_{u \in c} \frac{|\{ (u,v) | (u,v) \in E, v \notin c\}|}{\deg(u)}
$$ 

\textit{Flake out degree fraction} of a cluster $c$  is defined as 
$$
\text{flake\_ODF}(c) =  \frac{ \left| \{  u \in c ; |(u,v) \in E, v \in c | < \deg(u) \} \right|}{|c|}
$$

\subsection*{separability}
The \textit{separability} of a cluster $c$  is defined as 
$$
\text{separability}(c) = \frac{\text{sum\_intra\_weight}(c)}{\text{sum\_inter\_weight}(c)}
$$

\subsection*{coefficient}
For a cluster, similar to a graph, \textit{global clustering coefficient} 
and \textit{local clustering coefficient} can be defined on the induced subgraph of the cluster.

\subsection{Score Measures}
Score Measures evaluate performance by comparing clusters to its corresponding ground truth. Score measures are better choices to evaluate performance when ground truth is available (e.g. benchmark graphs)

\subsection*{mutual information (MI)}
Assume $A$, $B$ are two clusterings of a graph $G=(V,E)$, the mutual information between the two clusters is defined as 

$$
mi(A,B; G) = \sum_{i=1}^{|A|} \sum_{j=1}^{|B|} \frac {|A_i \cap B_j | }{|V|} \log \frac{|V| |A_i \cap B_j |}{ |A_i | | B_j |}
$$

\subsection*{normalized mutual info score (NMI)}
Assume $A$, $B$ are two clusterings of a graph $G=(V,E)$, the normalized mutual information between the two clusters is defined as 

$$nmi(A,B; G) = \frac{2 * mi(A,B; G)}{H(A)+H(B) } $$
where $H(.)$ is the entropy function. 

\subsection*{adjusted mutual information (ami)}
Assume $A$, $B$ are two clusterings of a graph $G=(V,E)$, the adjusted normalized mutual information between the two clusters is defined as 

$$ami(A,B; G) = \frac{mi(A,B; G) - E(mi(A,B; G)) } { \frac{H(A)+H(B)}{2}- E(mi(A,B; G))} $$
where $H(.)$ is the entropy function and $E(.)$ is the adjustment for mutual information \citep{vinh2010information}.

\subsection*{adjusted rand score (ARS)}
Rand index is defined as 
$$ RI(A,B) = \frac{TP+TN}{TP+TN+FP+FN}$$ where
TP is the number of true positives,  TN is the number of true negatives,  FP is the number of false positives, and FN is the number of false negatives.

Adjusted rand score is the corrected-for-chance version of the rand index
\citep{vinh2010information,hubert1985comparing} by using the expected similarity of all pair-wise comparisons between clusterings specified by a random model

\subsection*{V-measure score}
Support $A$ is the ground truth, and $C$ is a clustering, 
V-measure score \citep{rosenberg2007v} is the harmonic mean between homogeneity (or purification) and completeness, that is
 $$\text{V-measure} =  \frac {  homogeneity * completeness} {homogeneity + completeness} $$
where 
$ homogeneity(A,C) =  1 - \frac{H(A|C)}{H(A)}$
, $ completeness(A,C) =  1 - \frac{H(C|A)}{H(C)}$
where $H(.)$ is entropy function and $H(.|.)$ is conditional entropy function.

\section{Algorithms}

This section describes a few of algorithms that were tested in this paper. The descriptions are mainly excerpted from the original papers and not all the algorithms are covered. Please find references by Table~\ref{tbl:graph_support} or online documentation. 

\subsection*{Infomap}

Infomap \citep{Rosvall2008}  introduced an information theoretic approach that reveals community structure in weighted and directed graphs. To find a partition $M$ of a graph, the algorithm minimizes the expected description length of a random walk (a.k.a map equation) defined as 
$$ L(M) = q H(\mathcal{Q}) + \sum_{i=1}^m p^i H(\mathcal{P}^i) $$
where  the first part is the entropy of the movement between partitions, and the second part is the entropy of movements within partitions.

\subsection*{Hierarchical Infomap (infohiermap)}
Hierarchical map equation \citep{Rosvall2011}  extended map equation to enable multiple levels partitions of a graph. 
$$ L(M) = q H(\mathcal{Q}) + \sum_{i=1}^m p^i L(M^i)) $$
where $L(M^i)$ is the description length of submap $M^i$ which is defined recursively
until the finest level 
$$ L(M^{ij...}) =  \sum_{i=1}^m p^{ij...} L(M^{ij...})) $$

\subsection*{Order Statistics Local Optimization Method (OSLM)}
OSLOM \citep{Lancichinetti2011} is capable to detect clusters in networks accounting for edge directions, edge weights, overlapping communities, hierarchies and community dynamics. It is based on the local optimization of a fitness function expressing the statistical significance of clusters with respect to random fluctuations, which is estimated with tools of Extreme and Order Statistics.

\subsection*{COPRA (Community Overlap PRopagation Algorithm)}
COPRA \citep{Gregory2010} finds  overlapping community structure by  extending the label propagation algorithm to include information about more than one community.

\subsection*{Louvain}
Louvain \citep{Blondel2008} is method that  greedily  optimize the modularity of a graph. 
The algorithm initially assigns a partition for each vertex, then a  two-step process are repeated until  maximum modularity is achieved. 

In the first step it greedily merges partitions until no gain got.  In the second step, 
a new graph is built whose nodes are partitions in the first step. 

\subsection*{Label Propagation Method/Algorithm(LPM, LPA)}

LPA or LPM \citep{Raghavan2007} is a simple algorithm that propagates the labels (partition ids) to neighbors, then updates the labels  by voting. 

\subsection*{GANXiSw (SLPA)}
SLPA \citep{Xie2011} (Speaker-listener Label Propagation Algorithm) is an extension of LPA that discovers overlapping structures according to dynamic interaction rules

\subsection*{HiReCS}
HiReCS \footnote{http://www.lumais.com/hirecs}  a C++ clustering library for the multi-scale hierarchical community structure discovery with crisp overlaps.

\subsection*{LabelRank}
LabelRank \citep{xie2013labelrank} proposed strategies to stabilized the LPA and
to extend MCL approach to resolve the randomness issue in traditional label propagation algorithms (LPA).

\subsection*{CONGA}
CONGA \citep{Gregory2007} (Cluster-Overlap Newman Girvan Algorithm) is a method that discovers overlapping communities in networks, by extending Girvan and Newman’s well-known algorithm based on the betweenness centrality measure. 

\subsection*{CliquePercolation}
CliquePercolation \citep{Kumpula2008} is a fast community detection method in weighted and unweighted networks, for cliques of a chosen size. It is based on sequentially inserting the constituent links to the network and simultaneously keeping track of the emerging community structure.

\subsection*{Connected Iterative Scan(CIS)}
Connected Iterative Scan is also known at times as Locally Optimal Sets. Refer to \citep{Kelley2009} for details.
 
\subsection*{DEMON}
DEMON (Democratic Estimate of the Modular Organization of a Network) \citep{Coscia2012} is 
a  local-first approach to cluster discovery.
It first democratically let each node vote for the communities in its limited surrounding view of the global system using a label propagation algorithm. Then the local communities are merged into a global collection.

\subsection*{EAGLE}
EAGLE (agglomerativE hierarchicAl clusterinG based on maximaL cliquE) \citep{Shen2009} is a algorithm
that detects both the overlapping and hierarchical properties of complex community structure together. 
It deals with the set of maximal cliques and adopts an agglomerative framework, where the quality function of modularity is extended to evaluate the goodness of a cover.

\subsection*{FastCpm}
FastCpm \citep{Reid2012a} 
is  a simple algorithm to conduct clique percolation via the Bron Kerbosch algorithm. It claimed to perform much better than Sequential Clique Percolation (SCP) algorithm \citep{Kumpula2008}, especially for higher values of k.

\subsection*{Greedy Clique Expansion (GCE) Community Finder }

GCE \citep{Lee2010} first detects a set of seeds in graph G, then expanding these seeds in series by greedily maximizing a local community fitness function, and then finally accepts only those communities that are not near-duplicates of communities that have already been accepted. The fitness function of community $S$ is defined as 
$$
F_S=\frac{k_{in}^{S}}{(k_{in}^{S}+k_{out}^{S})^\alpha}
$$
where $k_{in}$ and $k_{out}$ is the internal and external degrees of $S$.

\subsection*{Hierarchical Demon (HDEMON)}
Hierarchical Demon \citep{Coscia2014} extended the DEMON algorithm by introducing the possibility of returning the hierarchical organization of the communities.

\subsection*{Link communities algorithm}
Link communities algorithm \citep{ahn2010link} is a hierarchical clustering algorithm through link dendrogram by a similarity function between links
$$
S(e_{ik},e_{jk}) = \frac{|n_+(i) \cap n_+(j)|}{|n_+(i) \cup n_+(j)|}
$$
where $n_+(i)$ is the the neighbor nodes of node $i$.

\subsection*{Model-based Overlapping Seed Expansion (MOSE)}
MOSE \citep{McDaid2010} algorithm is 
based on a statistical model of community structure, which is capable of detecting highly overlapping community structure, especially when there is variance in the number of communities each node is in.

\subsection*{Fast Multi-Scale Community Detection Tools (MSCD)}
MSCD \citep{LeMartelot2013} is a method compatible with global and
local criteria that enables fast multi-scale community detection on large networks.
It was implemented with 6 known criteria of HSLSW, LFK, RB, RN, SO and AFG. 

\subsection*{Model-based Overlapping Seed Expansion (ParCPM)}
ParCPM \citep{Gregori2013} is a novel, parallel k-clique community detection method, based on an innovative technique which enables connected components of a network to be obtained from those of its subnetworks. 

\subsection*{SVINET}
SVINET \citep{gopalan2013efficient} implements sampling based algorithms that derive from stochastic variational inference under the (assortative) mixed-membership stochastic blockmodel.

\subsection*{Top Graph Clusters (TopGC)}
TopGC \citep{Macropol2010} implements probabilistically finds the best well connected, clique-like clusters within large graphs. It is inherently parallelizable, and runs in linear time on the graph size.

\subsection*{Clique Modularity}
Clique modularity algorithm \citep{Yan2009} optimizes modularity by detecting disjoint cliques and then merges these.

\subsection*{CGGC (Core Groups Graph ensemble Clustering)}
CGGC \citep{Ovelgonne2013} is an ensemble learning algorithm to learn several weak classifiers and use these weak classifiers to determine a strong classifier.

\section{Benchmark Graphs}
We used several types of simulated graphs in our benchmark tests. 
Our focus is on undirected graphs since they are most commonly used graphs and are supported by most of algorithms.

\subsection{Lancichinetti-Fortunato-Radicchi (LFR) Benchmark }

LFR\citep{lancichinetti2008benchmark}
is an algorithm that generates benchmark networks with a priori known communities.
Overlapped and/or weighted LFR benchmark networks are described in \citep{lancichinetti2009benchmarks}.

LFR requires specifying at least four parameters $N$, $k$, $max_k$ and $\mu$ for unweighted graphs, where $N$ is the number of node, $k$ is the average in-degree, $max_k$ is the maximum in-degree and $\mu$ is a mixing parameter (which controls how clusters "mix" together"). For weighted graphs, weight mixing parameter $\mu_t$ is introduced. 

Symbols of LFR($N$, $k$, $max_k$, $\mu$) and WLFR($N$, $k$, $max_k$, $\mu$, $\mu_t$) is used for unweighted and weighted LFR graphs in the following sections.

\subsection{Random Graph}
A random graph itself does not has any structure, but it is a baseline and can be used to evaluate how algorithms scale with graph size.

We use RAND($n$,$m$) to denote a random graph where $n$ and $m$ are number of nodes and number of edges.

\subsection{SIMPLE Graph}
The performance of a clustering algorithm might not behavior evenly. It might  be affected by
\begin{itemize}
    \item graph size
    \item cluster sizes 
    \item inner degrees 
    \item inter degrees 
    \item other factors
\end{itemize}

While LFR is a widely used benchmark, it mixed these factors together so that it is hard to do analysis like performance attribution. 

We introduce a simple structural graph,  SIMPLE($nc$, $cz$, $k_i$, $k_o$), for our analysis where $nc$ is the number of cluster, $cz$ is the cluster size, $k_i$ is the internal degree for each node in a cluster and $k_o$ is  the number of inter-edges between two clusters. We also define $\gamma=k_o/k_i$ which measures how well a cluster is separated from its peers.

\section{Benchmark Result Analysis}
We collected 70+  graph clustering programs for our tests. These programs origin from sources of github, open source graph libraries or sharing by paper authors. 
An algorithm might be  implemented by different authors or be improved in some way for performance reason. 

To distinguish these programs we use name convenience of "\textit{prefix\_name}" where "prefix" is the source of the code and "name" is the algorithm name that used by the source which is easy for readers to identify it. 
A summary of prefixes can be found in Table~\ref{tbl:alg_prefix} and further details can be found in our code repository of \url{https://bitbucket.org/LizhenShi/graph_clustering_toolkit}.

\begin{table*}[t]
\centering
\caption{Description of algorithms prefixes}
\label{tbl:alg_prefix}

\begin{tabular}{|l|l|}
\hline
\textbf{Prefix} & \textbf{Description}                                        \\ \hline
mcl             & https://micans.org/mcl/                                     \\ \hline
igraph~\citep{igraph}          & https://igraph.org/                                         \\ \hline
sklearn~\citep{scikit-learn}         & https://scikit-learn.org/                                   \\ \hline
scan            & scan family  (scan, pscan, anyscan etc)                     \\ \hline
dct             & https://github.com/kit-algo/distributed\_clustering\_thrill \\ \hline
cgcc            & https://github.com/eXascaleInfolab/CGGC                     \\ \hline
networkit~\citep{staudt2016networkit}       & https://networkit.github.io/                                \\ \hline
snap~\citep{leskovec2016snap}            & https://snap.stanford.edu                                   \\ \hline
cdc             & https://github.com/RapidsAtHKUST/CommunityDetectionCodes    \\ \hline
oslom           & http://www.oslom.org/index.html                             \\ \hline
pycabem         & https://github.com/eXascaleInfolab/PyCABeM                  \\ \hline
karateclub~\citep{karateclub2020}      & https://github.com/benedekrozemberczki/karateclub           \\ \hline
alg             & others                                                      \\ \hline
\end{tabular}%
\end{table*}

\subsection{Graph Supports}

A graph could be  undirected/directed, weighted/unweighted and clustering  could also be unoverlapped or  overlapped. An algorithm usually does not  support all these types. Unfortunately  sources and papers may do not specifies its support clearly. We have to "guess" it by tests.

To determine the support of directed graph for an algorithm, a directed graph $G_{d}$ is made first, then is converted to its corresponding undirected graph $G_{ud}$. The algorithm is applied to both graphs and the clustering results are compared, where the algorithm is considered supporting directed graph if the results are not consistent. 
A similar approach is applied to determine the support of weighted graph.
However be aware that this method  might not be $100\%$ accurate.

The results are presented in Table~\ref{tbl:graph_support}, which shows of all the algorithms, $25\%$ support 
directed graph, $51\%$ support weighted graph and $11\%$ support overlapped graph.

\subsection{Runtime Benchmark}

Nowadays graph data is common to have number of edges up to millions and billions, so runtime is a big concern when choosing algorithms.
Despite the fact that most clustering algorithms are only capable for small or medium graph size,
we are still interested in which algorithms have lesser runtime cost, so that we can try them first or try to implement a parallel version when being necessary.

We benchmarked algorithms on three types of graphs (RAND, SIMPLE, LRF) where graph sizes (\#node) ranging from 32 to 100K.  The maximum runtime is limited up to 1 hour and the core number is limited to 1 to be fair. 

The results in Table~\ref{tbl:graph_runtime} shows the ranks that are sorted by the median of the three. Half of the algorithms can handle 100K node graph in 1 hour while the slowest igraph\_community\_optimal\_modularity can only handle hundreds of nodes. 
The actual maximum runtime ranges from 1 second to 1 hour and the top 30 finished in 100 seconds.
From the table we can see that most of the tops are based on LPM, scan and louvain methods. 

Be warned that the result is not highly accurate might because of the facts like 
\begin{itemize}
    \item code languages make big difference (c/c++, java, python)
    \item algorithm may  pre-process the graph into binary format 
    \item algorithm is designed for parallel processing, while it is limited to 1 core
\end{itemize}

\subsection{Clustering Quality Benchmark}
Clustering quality is critical for algorithms. We tested for the quality with both LFR and SIMPLE graphs. 
NMI is selected as criterion since ground truth is available (it was found that measures solely based 
on clusters have their drawbacks and have to be used with other measures together). Overlapped clustering was transferred to an unoverlapped one by randomly choosing one label for overlapped nodes when computing NMI.

\subsection{LFR benchmark}
\subsubsection*{Unweighted Graph}
We tested the graphs of $LFR(N,N/4,N/2,\mu)$ where $ N \in \{128, 512,1024\}$ and $\mu$ ranges from 0.1 to 0.9. 
Table~\ref{tbl:lfr_nmi_rank_short} shows top-rank algorithms
(full list in Table~\ref{tbl:lfr_nmi_rank}).
\begin{center}
\begin{adjustwidth}{-2cm}{}
\small 
\begin{longtable} {|l|l|l|l|l|l|l|l|l|l|l|}

\caption{NMI ranks for LFR weighted benchmark (short)}
\label{tbl:lfr_nmi_rank_short}\\

\hline 
\textbf{Algorithm}                             & \textbf{$\mu_{0.1}$} & \textbf{$\mu_{0.2}$} & \textbf{$\mu_{0.3}$} & \textbf{$\mu_{0.4}$} & \textbf{$\mu_{0.5}$} & \textbf{$\mu_{0.6}$} & \textbf{$\mu_{0.7}$} & \textbf{$\mu_{0.8}$} & \textbf{$\mu_{0.9}$} & \textbf{Median} \\
\endfirsthead
\multicolumn{11}{c}%
{\tablename\ \thetable{} -- continued from previous page} \\

\hline 
\textbf{Algorithm}                             & \textbf{$\mu_{0.1}$} & \textbf{$\mu_{0.2}$} & \textbf{$\mu_{0.3}$} & \textbf{$\mu_{0.4}$} & \textbf{$\mu_{0.5}$} & \textbf{$\mu_{0.6}$} & \textbf{$\mu_{0.7}$} & \textbf{$\mu_{0.8}$} & \textbf{$\mu_{0.9}$} & \textbf{Median} \\
\endhead

\hline \multicolumn{11}{|r|}{{Continued on next page}} \\ \hline
\endfoot

\hline
\endlastfoot

\hline
oslom\_OSLOM                            & 1                & 1                & 6                & 5                & 1                & 2                & 1                & 2                & 3                & 2               \\ \hline
igraph\_community\_spinglass            & 2                & 2                & 2                & 1                & 3                & 4                & 2                & 8                & 16               & 4               \\ \hline
dct\_dlplm                              & 1                & 1                & 1                & 3                & 4                & 5                & 5                & 13               & 16               & 5               \\ \hline
oslom\_modopt                           & 1                & 1                & 1                & 2                & 4                & 5                & 3                & 10               & 16               & 5               \\ \hline
igraph\_community\_walktrap             & 1                & 1                & 1                & 4                & 3                & 1                & 4                & 10               & 17               & 5               \\ \hline
igraph\_community\_leading\_eigenvector & 1                & 3                & 4                & 4                & 10               & 8                & 6                & 6                & 13               & 6               \\ \hline
dct\_dlslm\_no\_contraction             & 1                & 1                & 3                & 8                & 5                & 6                & 8                & 11               & 14               & 6               \\ \hline
oslom\_louvain\_method                  & 1                & 1                & 3                & 7                & 4                & 6                & 5                & 12               & 16               & 6               \\ \hline
dct\_dlslm                              & 1                & 1                & 1                & 5                & 5                & 5                & 3                & 13               & 16               & 6               \\ \hline
dct\_seq\_louvain                       & 1                & 1                & 1                & 7                & 2                & 5                & 7                & 12               & 16               & 6               \\ \hline
dct\_dlslm\_with\_seq                   & 1                & 1                & 1                & 4                & 6                & 4                & 5                & 13               & 16               & 6               \\ \hline
networkit\_PLM                          & 1                & 1                & 1                & 4                & 7                & 5                & 7                & 13               & 16               & 6               \\ \hline
igraph\_community\_multilevel           & 1                & 1                & 1                & 8                & 3                & 4                & 7                & 13               & 16               & 6               \\ \hline
karateclub\_EgoNetSplitter              & 8                & 4                & 2                & 6                & 6                & 4                & 4                & 10               & 15               & 7               \\ \hline
alg\_lso\_cluster                       & 1                & 1                & 5                & 10               & 9                & 7                & 11               & 14               & 15               & 8               \\ \hline
igraph\_community\_fastgreedy           & 1                & 3                & 6                & 9                & 8                & 8                & 10               & 11               & 16               & 8               \\ \hline
snap\_Clauset\_Newman\_Moore            & 1                & 5                & 8                & 12               & 12               & 9                & 9                & 6                & 14               & 8               \\ \hline
oslom\_Infohiermap                      & 1                & 1                & 13               & 22               & 18               & 10               & 2                & 1                & 2                & 8               \\ \hline
karateclub\_DANMF                       & 13               & 13               & 9                & 13               & 15               & 12               & 4                & 1                & 1                & 9               \\ \hline

\end{longtable}%
\end{adjustwidth}
\end{center}

 OSLOM is the best one which ranks top for all the $\mu$'s. Several algorithms from iGraph are also good. If being interested in large graphs, algorithms from dct family are worth trying since they support distributed computing. 

\begin{figure}
\centering
\includegraphics[scale=0.6]{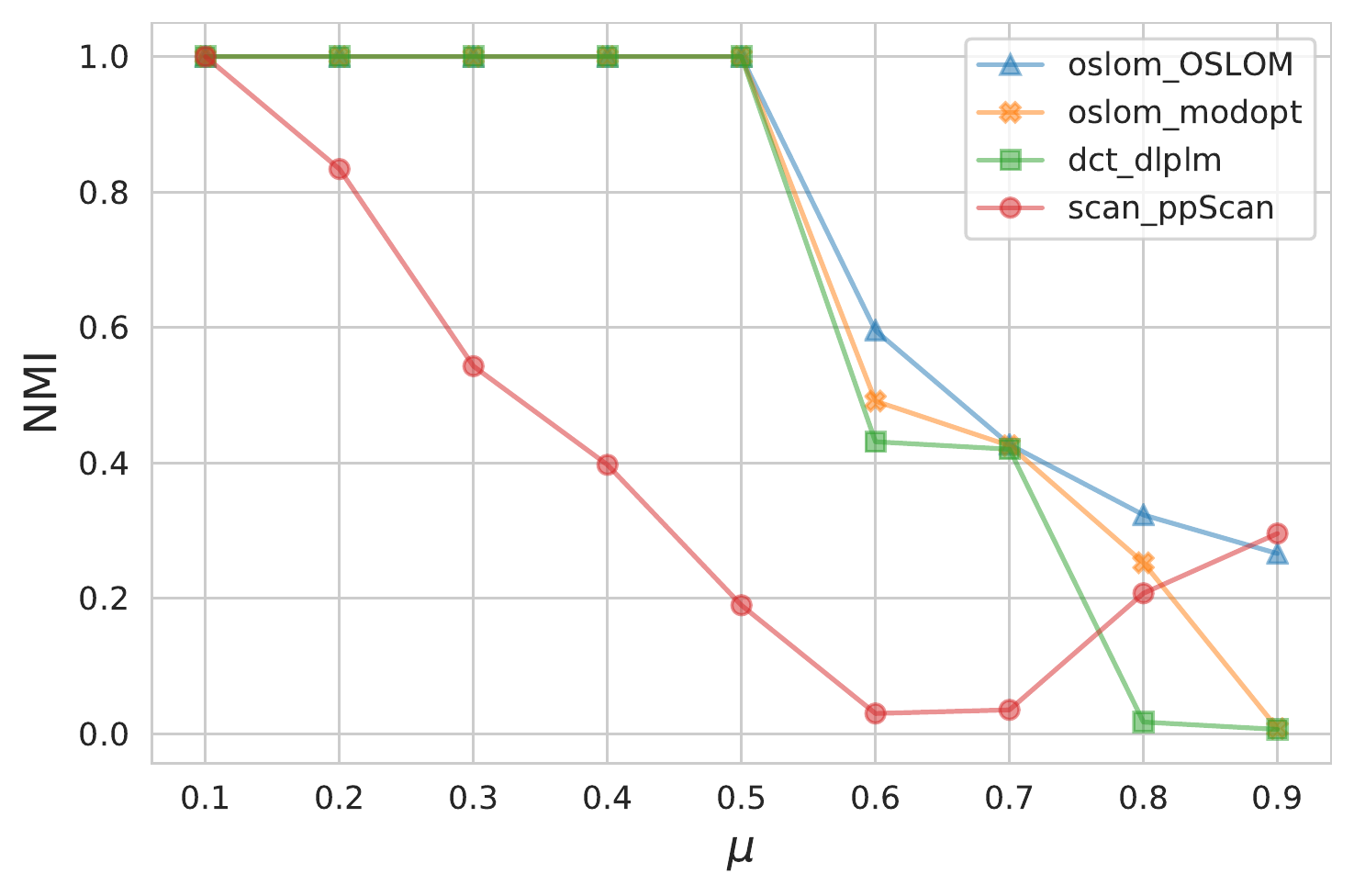}
\caption{ Selective Comparison of NMI Performance for LFR Benchmark }
\label{fig:lfr_nmi_linechart}
\end{figure}

Ranks does not reveal the whole story, actually NMI's for  all the algorithms drop quickly when $\mu$ goes beyond 0.5. Figure~\ref{fig:lfr_nmi_linechart} shows a few examples of 
algorithms based on OSLOM, modularity optimization, Louvain and scan. 

\subsubsection*{Weighted Graph}
\begin{figure}
\centering
\includegraphics[scale=0.6]{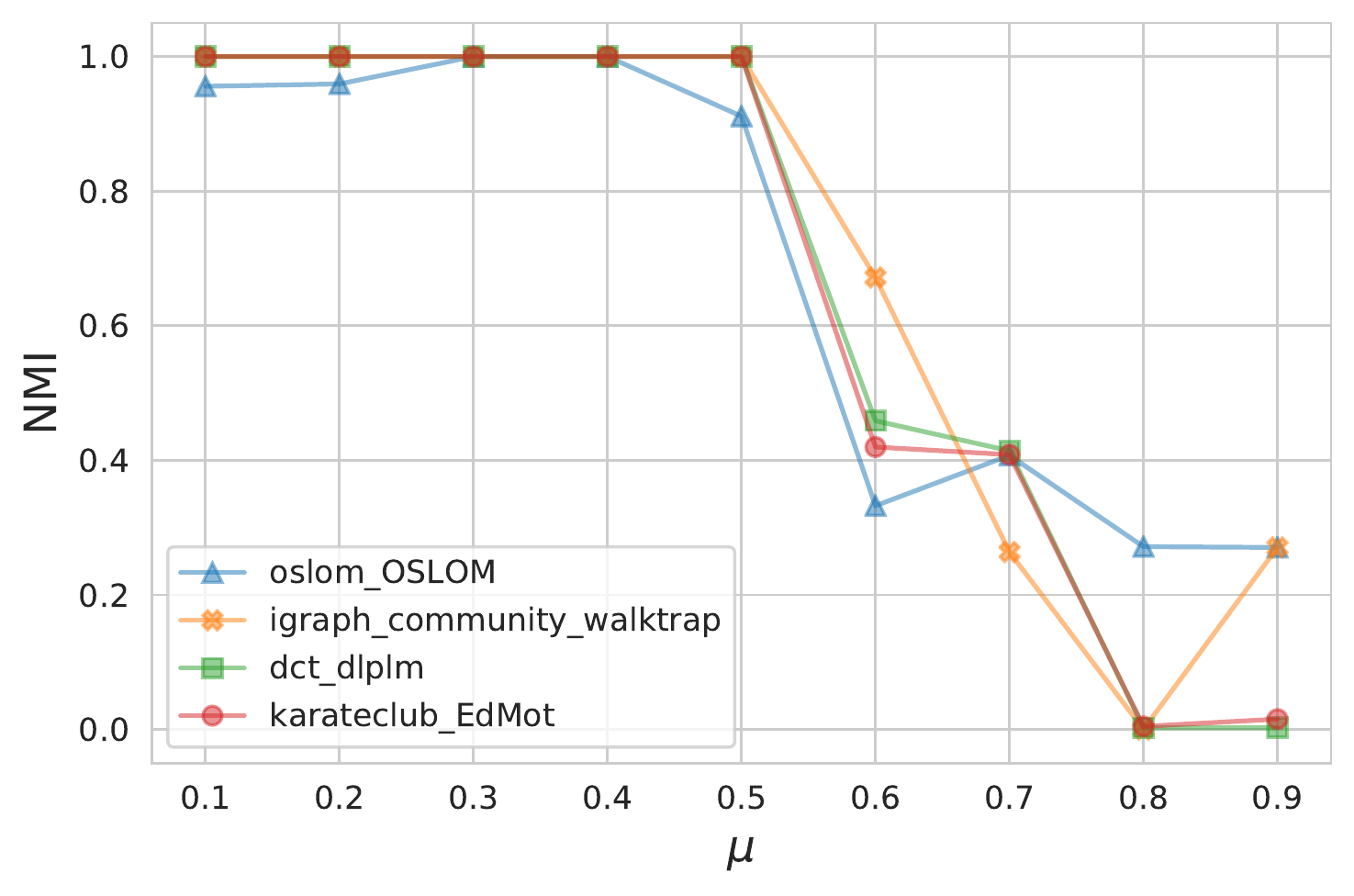}
\caption{ Selective Comparison of NMI Performance for LFR Weighted Benchmark }
\label{fig:lfr_nmi_w_linechart}
\end{figure}

We used similar configuration of $WLFR(N,N/4,N/2,\mu,\mu)$ where $ N \in \{128, 512,1024\}$ and $\mu$ ranges from 0.1 to 0.9. 
Table~\ref{tbl:lfr_w_nmi_rank_short} shows the top-rank algorithms
(full list in Table~\ref{tbl:lfr_w_nmi_rank}).
NMI's also drop quickly when $\mu$ is higher than 0.5. Figure~\ref{fig:lfr_nmi_w_linechart} shows a few examples.

\begin{table}[]
\caption{NMI ranks for LFR weighted benchmark (short)}
\label{tbl:lfr_w_nmi_rank_short}
\resizebox{\textwidth}{!}{%
\begin{tabular}{|l|l|l|l|l|l|l|l|l|l|l|}
\hline
\textbf{Algorithm}                      & \textbf{$\mu$\_0.1} & \textbf{$\mu$\_0.2} & \textbf{$\mu$\_0.3} & \textbf{$\mu$\_0.4} & \textbf{$\mu$\_0.5} & \textbf{$\mu$\_0.6} & \textbf{$\mu$\_0.7} & \textbf{$\mu$\_0.8} & \textbf{$\mu$\_0.9} & \textbf{Mean} \\ 

\hline
igraph\_community\_walktrap             & 1                           & 1                           & 2                           & 3                           & 1                           & 1                           & 1                           & 3                           & 5                           & 2                            \\ \hline
oslom\_OSLOM                            & 3                           & 2                           & 4                           & 4                           & 7                           & 6                           & 1                           & 1                           & 2                           & 3                            \\ \hline
oslom\_modopt                           & 1                           & 1                           & 3                           & 3                           & 2                           & 4                           & 2                           & 8                           & 12                          & 4                            \\ \hline
dct\_dlplm                              & 1                           & 1                           & 1                           & 1                           & 3                           & 2                           & 6                           & 8                           & 13                          & 4                            \\ \hline
dct\_dlslm                              & 1                           & 1                           & 2                           & 1                           & 3                           & 5                           & 5                           & 9                           & 13                          & 4                            \\ \hline
dct\_seq\_louvain                       & 1                           & 1                           & 1                           & 2                           & 3                           & 2                           & 7                           & 9                           & 13                          & 4                            \\ \hline
igraph\_community\_leading\_eigenvector & 2                           & 1                           & 1                           & 5                           & 9                           & 7                           & 3                           & 8                           & 7                           & 5                            \\ \hline
igraph\_community\_spinglass            & 2                           & 1                           & 5                           & 3                           & 5                           & 7                           & 5                           & 8                           & 12                          & 5                            \\ \hline
karateclub\_EgoNetSplitter              & 4                           & 3                           & 3                           & 1                           & 3                           & 7                           & 7                           & 9                           & 12                          & 5                            \\ \hline
igraph\_community\_multilevel           & 1                           & 1                           & 6                           & 3                           & 3                           & 3                           & 4                           & 9                           & 12                          & 5                            \\ \hline
networkit\_PLM                          & 1                           & 1                           & 6                           & 3                           & 3                           & 3                           & 4                           & 9                           & 12                          & 5                            \\ \hline
alg\_lso\_cluster                       & 1                           & 1                           & 6                           & 3                           & 2                           & 5                           & 6                           & 7                           & 11                          & 5                            \\ \hline
karateclub\_EdMot                       & 1                           & 4                           & 6                           & 3                           & 2                           & 5                           & 4                           & 9                           & 12                          & 5                            \\ \hline
oslom\_Infohiermap                      & 1                           & 5                           & 8                           & 10                          & 12                          & 9                           & 1                           & 1                           & 2                           & 5                            \\ \hline
oslom\_louvain\_method                  & 1                           & 4                           & 6                           & 3                           & 4                           & 8                           & 6                           & 9                           & 11                          & 6                            \\ \hline
igraph\_community\_fastgreedy           & 5                           & 6                           & 7                           & 4                           & 6                           & 6                           & 6                           & 8                           & 12                          & 7                            \\ \hline
karateclub\_SymmNMF                     & 14                          & 9                           & 11                          & 7                           & 10                          & 12                          & 3                           & 2                           & 1                           & 8                            \\ \hline
sklearn\_SpectralClustering             & 12                          & 9                           & 9                           & 6                           & 8                           & 11                          & 9                           & 7                           & 12                          & 9                            \\ \hline

\end{tabular}%
}
\end{table}

\subsection{SIMPLE benchmark}

SIMPLE benchmark was performed with graphs of 1024 nodes while $cz$ (cluster size) varies from 4 to 256, and with a set of ($k_i$, $k_o$). 

\subsubsection*{Cluster Size Effect}
Table~\ref{tbl:simple_cz_nmi_rank_short} (full in Table~\ref{tbl:simple_cz_nmi_rank}) shows the NMI ranking by cluster size. 
We see that OSLOM performs the best and Louvain based algorithms are also in the tops. Another observation is that some algorithms behave better for larger cluster size, while a few of them behave better for smaller graph size.

\begin{figure}
\centering
\includegraphics[scale=0.6]{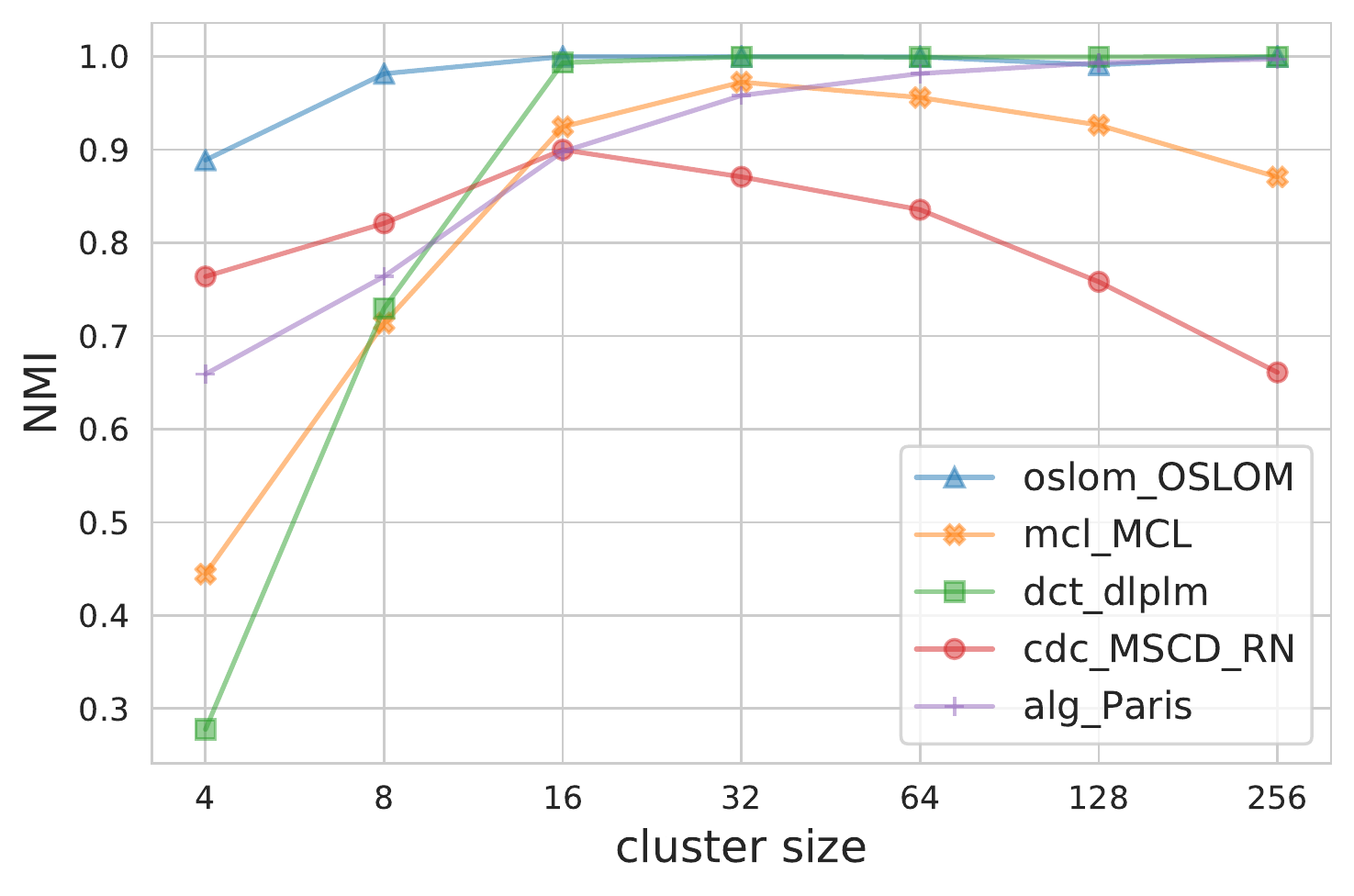}
\caption{ Selective Comparison of NMI Performance by cluster size for SIMPLE Benchmark }
\label{fig:simple_nmi_cz_linechart}
\end{figure}

Figure~\ref{fig:simple_nmi_cz_linechart} shows examples how NMI changes along with cluster sizes.

\begin{center}
\begin{adjustwidth}{-2cm}{}
\small 
\begin{longtable}{|l|l|l|l|l|l|l|l|l|}

\caption{NMI ranks by cluster size for SIMPLE benchmark (short)}
\label{tbl:simple_cz_nmi_rank_short}\\

\hline 
\textbf{Algorithm}                                            & \textbf{cz004} & \textbf{cz008} & \textbf{cz016} & \textbf{cz032} & \textbf{cz064} & \textbf{cz128} & \textbf{cz256} & \textbf{Mean} \\ \hline
\endfirsthead
\multicolumn{9}{c}%
{\tablename\ \thetable{} -- continued from previous page} \\

\hline 
\textbf{Algorithm}                                        & \textbf{cz004} & \textbf{cz008} & \textbf{cz016} & \textbf{cz032} & \textbf{cz064} & \textbf{cz128} & \textbf{cz256} & \textbf{Mean} \\ \hline

\endhead

\hline \multicolumn{9}{|r|}{{Continued on next page}} \\ \hline
\endfoot

\hline
\endlastfoot

\hline
oslom\_OSLOM                            & 1              & 2              & 1              & 1              & 1              & 2              & 1              & 1             \\ \hline
oslom\_Infohiermap                      & 1              & 3              & 1              & 1              & 1              & 3              & 2              & 2             \\ \hline
alg\_Paris                              & 4              & 7              & 9              & 5              & 3              & 2              & 1              & 4             \\ \hline
dct\_dlplm                              & 13             & 9              & 2              & 1              & 1              & 1              & 1              & 4             \\ \hline
dct\_dlslm\_with\_seq                   & 15             & 16             & 2              & 1              & 1              & 1              & 1              & 5             \\ \hline
karateclub\_EdMot                       & 14             & 14             & 2              & 1              & 1              & 1              & 1              & 5             \\ \hline
dct\_seq\_louvain                       & 13             & 14             & 2              & 1              & 1              & 1              & 1              & 5             \\ \hline
dct\_dlslm                              & 13             & 15             & 2              & 1              & 1              & 1              & 1              & 5             \\ \hline
igraph\_community\_walktrap             & 19             & 11             & 1              & 1              & 1              & 1              & 1              & 5             \\ \hline
dct\_dlslm\_no\_contraction             & 10             & 12             & 2              & 2              & 3              & 4              & 3              & 5             \\ \hline
oslom\_louvain\_method                  & 12             & 8              & 2              & 2              & 3              & 3              & 3              & 5             \\ \hline
networkit\_PLM                          & 16             & 18             & 4              & 1              & 1              & 1              & 1              & 6             \\ \hline
cdc\_MSCD\_RB                           & 8              & 11             & 5              & 5              & 4              & 3              & 3              & 6             \\ \hline
igraph\_community\_multilevel           & 15             & 18             & 4              & 1              & 1              & 1              & 1              & 6             \\ \hline
cdc\_MOSES                              & 28             & 1              & 1              & 2              & 2              & 5              & 9              & 7             \\ \hline
cdc\_MSCD\_AFG                          & 12             & 17             & 5              & 5              & 4              & 3              & 3              & 7             \\ \hline
cdc\_MSCD\_SO                           & 13             & 15             & 5              & 5              & 4              & 3              & 3              & 7             \\ \hline
cdc\_MSCD\_SOM                          & 15             & 13             & 5              & 5              & 4              & 3              & 3              & 7             \\ \hline
oslom\_modopt                           & 21             & 27             & 2              & 2              & 2              & 1              & 1              & 8             \\ \hline
cdc\_GCE                                & 29             & 4              & 1              & 1              & 5              & 7              & 7              & 8             \\ \hline
mcl\_MCL                                & 7              & 10             & 7              & 4              & 5              & 8              & 12             & 8             \\ \hline
pycabem\_LabelRank                      & 1              & 8              & 14             & 2              & 4              & 10             & 19             & 8             \\ \hline
igraph\_community\_spinglass            & 18             & 22             & 13             & 6              & 2              & 1              & 1              & 9             \\ \hline

\end{longtable}%
\end{adjustwidth}
\end{center}

\subsubsection*{Gamma Effect}

Recall $\gamma = k_o/k_i$ which measures how well a cluster is separated from others. 
Since $\gamma$ is a continuous number, it was categorized into 5 bins from 0 to 1.

\begin{figure}
\begin{adjustwidth}{-1cm}{}
\centering
\begin{subfigure}{}
  \centering
  \includegraphics[width=.45\linewidth]{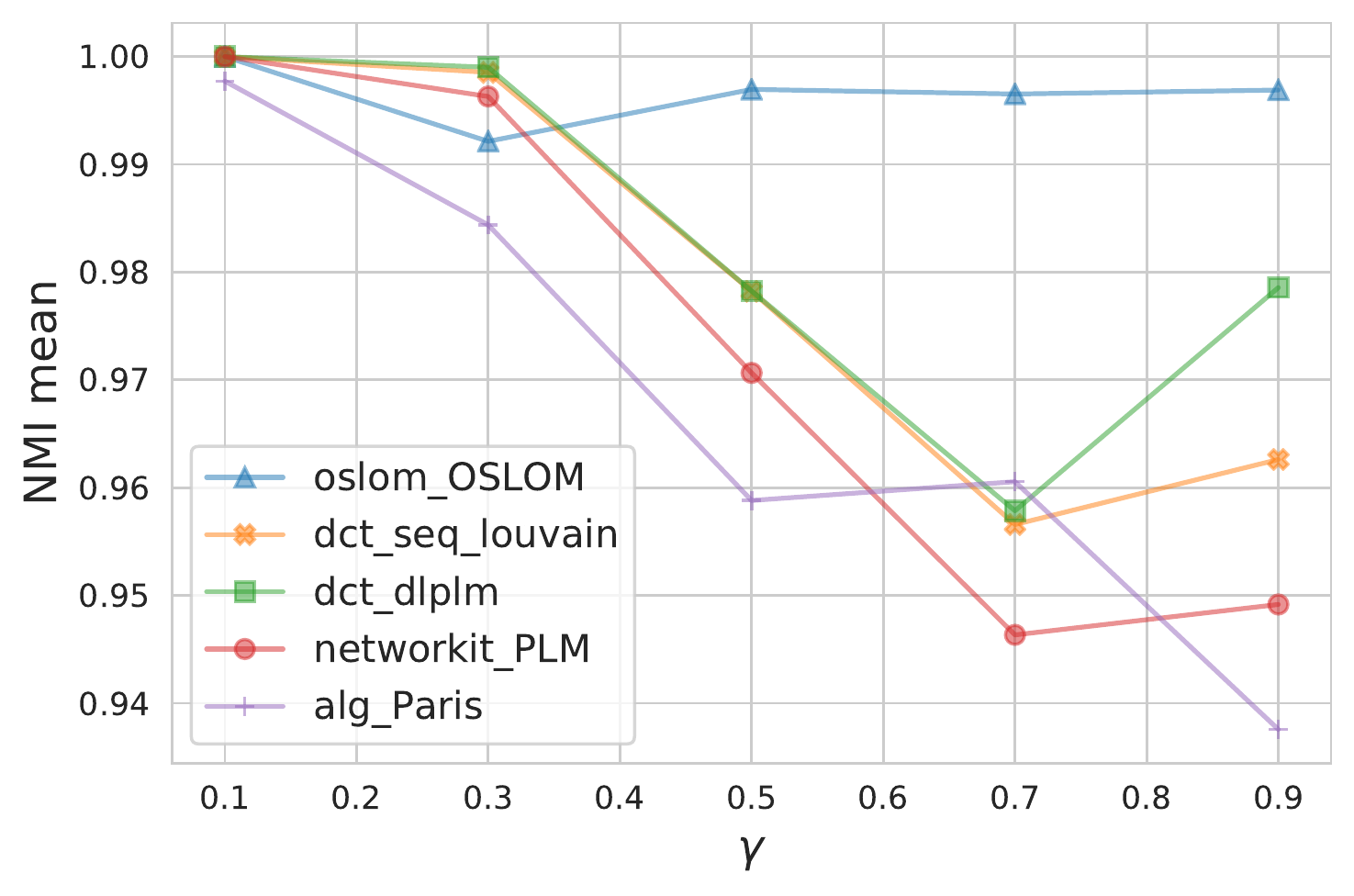}
\end{subfigure}%
\begin{subfigure}{}
  \centering
  \includegraphics[width=.45\linewidth]{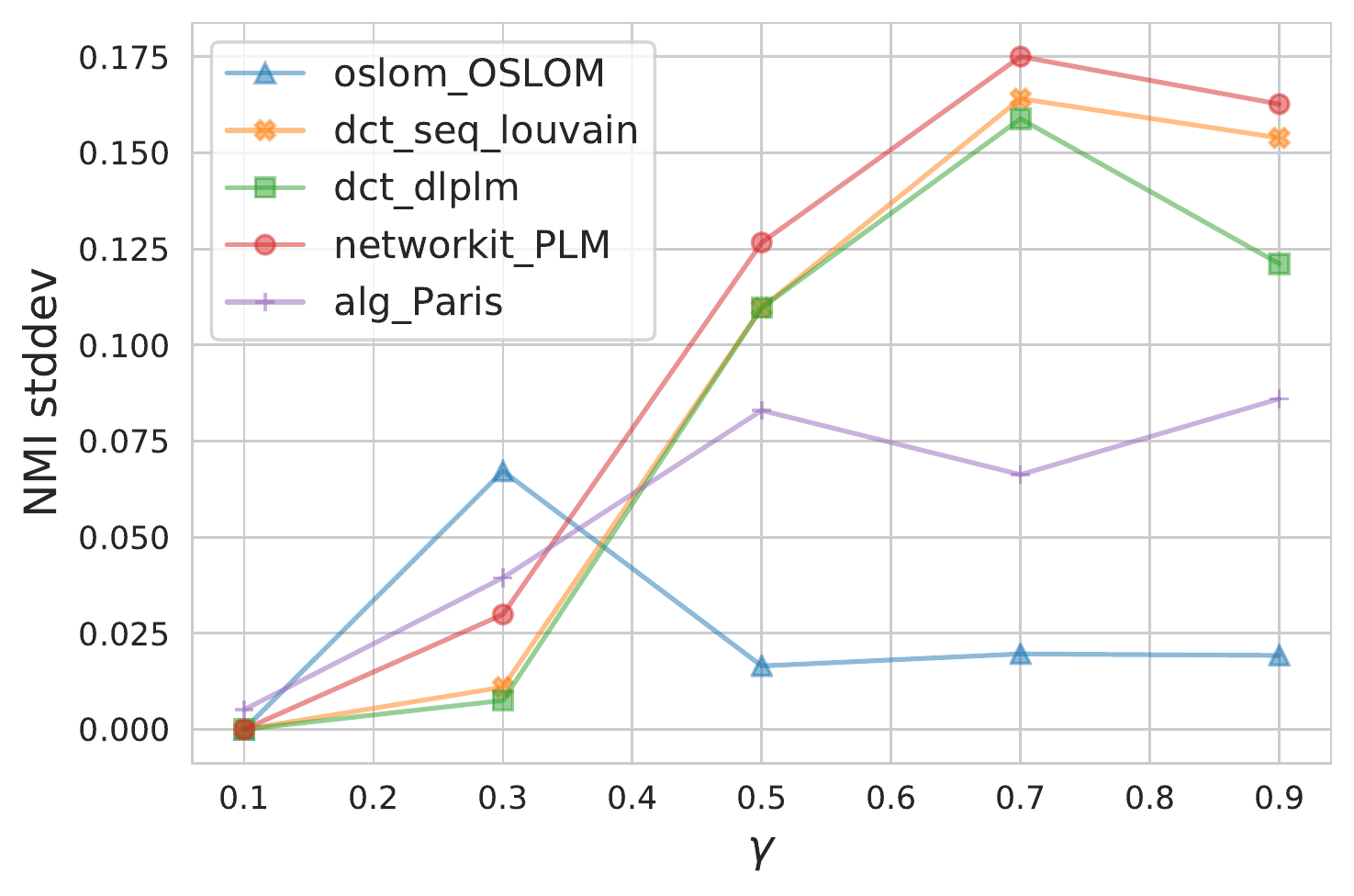}
\end{subfigure}
\caption{ Selective Comparison of NMI Performance by $\gamma$ for SIMPLE Benchmark }
\label{fig:simple_nmi_gamma_linechart}
\end{adjustwidth}
\end{figure}

Table~\ref{tbl:simple_gamma_nmi_rank_short} (full in Table~\ref{tbl:simple_gamma_nmi_rank}) shows the ranks for the bins and Figure~\ref{fig:simple_nmi_gamma_linechart} displays the actual changes.   
OSOM still outperforms others. 

\begin{center}
\begin{adjustwidth}{-4cm}{}
\small
\begin{longtable}{|l|c|c|c|c|c|c|c|}

\caption{NMI ranks by $\gamma$ for SIMPLE benchmark (short)}
\label{tbl:simple_gamma_nmi_rank_short}\\

\hline 
\textbf{Algorithm}                      & \textbf{$\gamma$(0.0-0.2)} & \textbf{$\gamma$(0.2-0.4)} & \textbf{$\gamma$(0.4-0.6)} &  \textbf{$\gamma$(0.6-0.8)} & \textbf{$\gamma$(0.8-1.0)} & \textbf{Mean}\\
\hline
\endfirsthead
\multicolumn{7}{c}%
{\tablename\ \thetable{} -- continued from previous page} \\

\hline 
\textbf{Algorithm}                      & \textbf{$\gamma$(0.0-0.2)} & \textbf{$\gamma$(0.2-0.4)} & \textbf{$\gamma$(0.4-0.6)} &  \textbf{$\gamma$(0.6-0.8)} & \textbf{$\gamma$(0.8-1.0)} & \textbf{Mean}\\
 \hline

\endhead

\hline \multicolumn{7}{|r|}{{Continued on next page}} \\ \hline
\endfoot

\hline
\endlastfoot

\hline
oslom\_OSLOM                            & 1                               & 2                               & 1                               & 1                               & 1                               & 1    \\ \hline
igraph\_community\_walktrap             & 1                               & 1                               & 2                               & 3                               & 3                               & 2    \\ \hline
dct\_dlplm                              & 1                               & 1                               & 2                               & 3                               & 2                               & 2    \\ \hline
karateclub\_EdMot                       & 1                               & 1                               & 2                               & 3                               & 4                               & 2    \\ \hline
dct\_seq\_louvain                       & 1                               & 1                               & 2                               & 3                               & 4                               & 2    \\ \hline
dct\_dlslm                              & 1                               & 1                               & 2                               & 3                               & 4                               & 2    \\ \hline
dct\_dlslm\_with\_seq                   & 1                               & 1                               & 2                               & 4                               & 4                               & 2    \\ \hline
oslom\_Infohiermap                      & 1                               & 2                               & 2                               & 2                               & 2                               & 2    \\ \hline
igraph\_community\_multilevel           & 1                               & 1                               & 3                               & 4                               & 5                               & 3    \\ \hline
networkit\_PLM                          & 1                               & 1                               & 3                               & 4                               & 5                               & 3    \\ \hline
dct\_dlslm\_map\_eq                     & 1                               & 2                               & 4                               & 5                               & 5                               & 3    \\ \hline
oslom\_modopt                           & 1                               & 3                               & 4                               & 4                               & 5                               & 3    \\ \hline
alg\_Paris                              & 1                               & 3                               & 4                               & 3                               & 6                               & 3    \\ \hline
dct\_infomap                            & 1                               & 3                               & 4                               & 5                               & 8                               & 4    \\ \hline
dct\_dlslm\_no\_contraction             & 1                               & 3                               & 6                               & 6                               & 6                               & 4    \\ \hline
oslom\_louvain\_method                  & 1                               & 2                               & 6                               & 5                               & 5                               & 4    \\ \hline
igraph\_community\_infomap              & 1                               & 2                               & 4                               & 5                               & 8                               & 4    \\ \hline
oslom\_lpm                              & 1                               & 3                               & 4                               & 6                               & 8                               & 4    \\ \hline
cdc\_MSCD\_AFG                          & 3                               & 4                               & 5                               & 6                               & 7                               & 5    \\ \hline
cdc\_MSCD\_RB                           & 3                               & 4                               & 5                               & 5                               & 6                               & 5    \\ \hline
igraph\_community\_spinglass            & 2                               & 4                               & 6                               & 6                               & 7                               & 5    \\ \hline
cdc\_MSCD\_HSLSW                        & 3                               & 4                               & 6                               & 6                               & 8                               & 5    \\ \hline
cdc\_MSCD\_SO                           & 3                               & 4                               & 5                               & 5                               & 6                               & 5    \\ \hline
oslom\_copra                            & 1                               & 4                               & 5                               & 6                               & 11                              & 5    \\ \hline
igraph\_community\_label\_propagation   & 1                               & 4                               & 4                               & 7                               & 8                               & 5    \\ \hline
cdc\_MSCD\_SOM                          & 3                               & 4                               & 5                               & 5                               & 6                               & 5    \\ \hline

\end{longtable}%
\end{adjustwidth}
\end{center}

\section{Summary}
In this paper we benchmarked more than 70 graph clustering programs for unweighted and weighted graphs. We tested their runtimes up to 100K nodes . When evaluating the quality of clustering, we focused on NMI metric and tested with LFR and SIMPLE benchmarks. Specifically we introduced SIMPLE benchmark to investigate how performance varies with cluster size, intra-degree and inter-degree.
Our work is able to not only supply a start point for engineers to select clustering algorithms but also could provide a viewpoint for researchers to design new  algorithms. 

We opened the source code which is shared at \url{https://bitbucket.org/LizhenShi/graph_clustering_toolkit} where documents and docker image are also provided. 

%
\clearpage 
\bibliographystyle{ACM-Reference-Format}
\bibliography{library.bib}

%
\clearpage
\appendix

\section{Full Tables of Benchmarks}

\begin{center}

\begin{longtable}{lccc}
\caption{Graph Support of Algorithms}
\label{tbl:graph_support}\\
\toprule

\hline 
\textbf{Algorithm}                            & \textbf{Directed?} & \textbf{ Weighted?} & \textbf{Overlapped?} \\
\endfirsthead
\multicolumn{4}{c}%
{\tablename\ \thetable{} -- continued from previous page} \\

\hline 
\textbf{Algorithm}                            & \textbf{Directed?} & \textbf{ Weighted?} & \textbf{Overlapped?} \\
\endhead

\hline \multicolumn{4}{|r|}{{Continued on next page}} \\ \hline
\endfoot

\hline
\endlastfoot

\hline 
oslom\_Infohiermap~\citep{Rosvall2011}                      & Y                          & Y                          & N                            \\ \hline
oslom\_Infomap~\citep{Rosvall2008}                          & N                          & Y                          & N                            \\ \hline
oslom\_OSLOM~\citep{Lancichinetti2011}                            & N                          & Y                          & Y                            \\ \hline
oslom\_copra~\citep{Gregory2010}                            & Y                          & Y                          & Y                            \\ \hline
oslom\_louvain\_method~\citep{Blondel2008}                  & N                          & Y                          & N                            \\ \hline
oslom\_lpm~\citep{Raghavan2007}                              & N                          & Y                          & N                            \\ \hline
oslom\_modopt~\citep{sales2007extracting}                           & N                          & Y                          & N                            \\ \hline
pycabem\_GANXiSw~\citep{Xie2011}                        & Y                          & Y                          & Y                            \\
\hline
\end{longtable}%
\end{center}

\begin{center}

\begin{longtable}{lcccc}
\caption{Runtime Benchmark Ranks}
\label{tbl:graph_runtime}\\
\toprule

\hline 
\textbf{Algorithm}                      & \textbf{RAND} & \textbf{SIMPLE} & \textbf{LFR} & \textbf{Median} \\
\endfirsthead
\multicolumn{5}{c}%
{\tablename\ \thetable{} -- continued from previous page} \\

\hline 
\textbf{Algorithm}                      & \textbf{RAND} & \textbf{SIMPLE} & \textbf{LFR} & \textbf{Median} \\
\endhead

\hline \multicolumn{5}{|r|}{{Continued on next page}} \\ \hline
\endfoot

\hline
\endlastfoot

 \hline
scan\_ppScanSSE                         & 1                  & 1                    & 8                 & 1                    \\ \hline
scan\_ppScan                            & 2                  & 2                    & 9                 & 2                    \\ \hline
networkit\_PLP                          & 3                  & 4                    & 3                 & 3                    \\ \hline
igraph\_community\_label\_propagation   & 12                 & 3                    & 2                 & 3                    \\ \hline
alg\_streamcom                          & 4                  & 5                    & 5                 & 5                    \\ \hline
networkit\_LPDegreeOrdered              & 69                 & 6                    & 4                 & 6                    \\ \hline
scan\_pScan                             & 5                  & 7                    & 28                & 7                    \\ \hline
igraph\_community\_multilevel           & 41                 & 8                    & 7                 & 8                    \\
\hline
\end{longtable}%
\end{center}

\begin{center}
\begin{adjustwidth}{-2cm}{}
\small 
\begin{longtable} {|l|l|l|l|l|l|l|l|l|l|l|}

\caption{NMI ranks for LFR unweighted benchmark (full)}
\label{tbl:lfr_nmi_rank}\\

\hline 
\textbf{Algorithm}                             & \textbf{$\mu_{0.1}$} & \textbf{$\mu_{0.2}$} & \textbf{$\mu_{0.3}$} & \textbf{$\mu_{0.4}$} & \textbf{$\mu_{0.5}$} & \textbf{$\mu_{0.6}$} & \textbf{$\mu_{0.7}$} & \textbf{$\mu_{0.8}$} & \textbf{$\mu_{0.9}$} & \textbf{Median} \\
\endfirsthead
\multicolumn{11}{c}%
{\tablename\ \thetable{} -- continued from previous page} \\

\hline 
\textbf{Algorithm}                             & \textbf{$\mu_{0.1}$} & \textbf{$\mu_{0.2}$} & \textbf{$\mu_{0.3}$} & \textbf{$\mu_{0.4}$} & \textbf{$\mu_{0.5}$} & \textbf{$\mu_{0.6}$} & \textbf{$\mu_{0.7}$} & \textbf{$\mu_{0.8}$} & \textbf{$\mu_{0.9}$} & \textbf{Median} \\
\endhead

\hline \multicolumn{11}{|r|}{{Continued on next page}} \\ \hline
\endfoot

\hline
\endlastfoot

\hline
oslom\_OSLOM                            & 1                & 1                & 6                & 5                & 1                & 2                & 1                & 2                & 3                & 2               \\ \hline
igraph\_community\_spinglass            & 2                & 2                & 2                & 1                & 3                & 4                & 2                & 8                & 16               & 4               \\ \hline
dct\_dlplm                              & 1                & 1                & 1                & 3                & 4                & 5                & 5                & 13               & 16               & 5               \\ \hline
oslom\_modopt                           & 1                & 1                & 1                & 2                & 4                & 5                & 3                & 10               & 16               & 5               \\ \hline
igraph\_community\_walktrap             & 1                & 1                & 1                & 4                & 3                & 1                & 4                & 10               & 17               & 5               \\ \hline
igraph\_community\_leading\_eigenvector & 1                & 3                & 4                & 4                & 10               & 8                & 6                & 6                & 13               & 6               \\ \hline
dct\_dlslm\_no\_contraction             & 1                & 1                & 3                & 8                & 5                & 6                & 8                & 11               & 14               & 6               \\ \hline
oslom\_louvain\_method                  & 1                & 1                & 3                & 7                & 4                & 6                & 5                & 12               & 16               & 6               \\
\hline
\end{longtable}%
\end{adjustwidth}
\end{center}

\begin{center}
\begin{adjustwidth}{-2cm}{}
\small 
\begin{longtable}{|l|l|l|l|l|l|l|l|l|}

\caption{NMI ranks by cluster size for SIMPLE benchmark}
\label{tbl:simple_cz_nmi_rank}\\

\hline 
\textbf{Algorithm}                                            & \textbf{cz004} & \textbf{cz008} & \textbf{cz016} & \textbf{cz032} & \textbf{cz064} & \textbf{cz128} & \textbf{cz256} & \textbf{Mean} \\ \hline
\endfirsthead
\multicolumn{9}{c}%
{\tablename\ \thetable{} -- continued from previous page} \\

\hline 
\textbf{Algorithm}                                        & \textbf{cz004} & \textbf{cz008} & \textbf{cz016} & \textbf{cz032} & \textbf{cz064} & \textbf{cz128} & \textbf{cz256} & \textbf{Mean} \\ \hline

\endhead

\hline \multicolumn{9}{|r|}{{Continued on next page}} \\ \hline
\endfoot

\hline
\endlastfoot

\hline
oslom\_OSLOM                            & 1              & 2              & 1              & 1              & 1              & 2              & 1              & 1             \\ \hline
oslom\_Infohiermap                      & 1              & 3              & 1              & 1              & 1              & 3              & 2              & 2             \\ \hline
alg\_Paris                              & 4              & 7              & 9              & 5              & 3              & 2              & 1              & 4             \\ \hline
dct\_dlplm                              & 13             & 9              & 2              & 1              & 1              & 1              & 1              & 4             \\ \hline
dct\_dlslm\_with\_seq                   & 15             & 16             & 2              & 1              & 1              & 1              & 1              & 5             \\ \hline
karateclub\_EdMot                       & 14             & 14             & 2              & 1              & 1              & 1              & 1              & 5             \\ \hline
dct\_seq\_louvain                       & 13             & 14             & 2              & 1              & 1              & 1              & 1              & 5             \\ \hline
dct\_dlslm                              & 13             & 15             & 2              & 1              & 1              & 1              & 1              & 5             \\
\hline

\end{longtable}%
\end{adjustwidth}
\end{center}

\begin{center}
\begin{adjustwidth}{-4cm}{}
\small
\begin{longtable}{|l|c|c|c|c|c|c|c|}

\caption{NMI ranks by $\gamma$ for SIMPLE benchmark}
\label{tbl:simple_gamma_nmi_rank}\\

\hline 
\textbf{Algorithm}                      & \textbf{$\gamma$(0.0-0.2)} & \textbf{$\gamma$(0.2-0.4)} & \textbf{$\gamma$(0.4-0.6)} &  \textbf{$\gamma$(0.6-0.8)} & \textbf{$\gamma$(0.8-1.0)} & \textbf{Mean}\\
\hline
\endfirsthead
\multicolumn{7}{c}%
{\tablename\ \thetable{} -- continued from previous page} \\

\hline 
\textbf{Algorithm}                      & \textbf{$\gamma$(0.0-0.2)} & \textbf{$\gamma$(0.2-0.4)} & \textbf{$\gamma$(0.4-0.6)} &  \textbf{$\gamma$(0.6-0.8)} & \textbf{$\gamma$(0.8-1.0)} & \textbf{Mean}\\
 \hline

\endhead

\hline \multicolumn{7}{|r|}{{Continued on next page}} \\ \hline
\endfoot

\hline
\endlastfoot

\hline
oslom\_OSLOM                            & 1                               & 2                               & 1                               & 1                               & 1                               & 1    \\ \hline
igraph\_community\_walktrap             & 1                               & 1                               & 2                               & 3                               & 3                               & 2    \\ \hline
dct\_dlplm                              & 1                               & 1                               & 2                               & 3                               & 2                               & 2    \\ \hline
karateclub\_EdMot                       & 1                               & 1                               & 2                               & 3                               & 4                               & 2    \\ \hline
dct\_seq\_louvain                       & 1                               & 1                               & 2                               & 3                               & 4                               & 2    \\ \hline
dct\_dlslm                              & 1                               & 1                               & 2                               & 3                               & 4                               & 2    \\ \hline
dct\_dlslm\_with\_seq                   & 1                               & 1                               & 2                               & 4                               & 4                               & 2    \\ \hline
oslom\_Infohiermap                      & 1                               & 2                               & 2                               & 2                               & 2                               & 2    \\
\hline

\end{longtable}%
\end{adjustwidth}
\end{center}

\begin{table}[]
\caption{NMI ranks for LFR weighted benchmark}
\label{tbl:lfr_w_nmi_rank}
\resizebox{\textwidth}{!}{%
\begin{tabular}{|l|l|l|l|l|l|l|l|l|l|l|}
\hline
\textbf{Algorithm}                      & \textbf{$\mu$\_0.1} & \textbf{$\mu$\_0.2} & \textbf{$\mu$\_0.3} & \textbf{$\mu$\_0.4} & \textbf{$\mu$\_0.5} & \textbf{$\mu$\_0.6} & \textbf{$\mu$\_0.7} & \textbf{$\mu$\_0.8} & \textbf{$\mu$\_0.9} & \textbf{Mean} \\ 

\hline
igraph\_community\_walktrap             & 1                           & 1                           & 2                           & 3                           & 1                           & 1                           & 1                           & 3                           & 5                           & 2                            \\ \hline
oslom\_OSLOM                            & 3                           & 2                           & 4                           & 4                           & 7                           & 6                           & 1                           & 1                           & 2                           & 3                            \\ \hline
oslom\_modopt                           & 1                           & 1                           & 3                           & 3                           & 2                           & 4                           & 2                           & 8                           & 12                          & 4                            \\ \hline
dct\_dlplm                              & 1                           & 1                           & 1                           & 1                           & 3                           & 2                           & 6                           & 8                           & 13                          & 4                            \\
\hline
\end{tabular}%
}
\end{table}

\end{document}